\documentclass[aps,prd,reprint,showpacs,floatfix,a4paper, nofootinbib]{revtex4-1}

\usepackage[english]{babel}
\usepackage{graphicx}
\usepackage{array}
\usepackage{dcolumn}
\usepackage{bm}
\usepackage{amssymb}
\usepackage{amsmath}
\usepackage{amsfonts}
\usepackage{amsbsy}
\usepackage{hyperref}

\begin{document}

\title{Exact theory of freeze out }

\author{Mirco Cannoni}
\affiliation{Departamento de F\'isica Aplicada, Facultad de Ciencias
Experimentales, Universidad de Huelva, 21071 Huelva, Spain}

\begin{abstract}

We show that the standard theory of thermal production and chemical decoupling of WIMPs is incomplete.
The hypothesis that WIMPs are produced and decouple from a thermal bath implies that the 
rate equation the bath particles interacting with the WIMPs is an algebraic equation that constraints 
the actual WIMPs abundance to have a precise analytical form down to the temperature $x_\ast=m_\chi /T_\ast$. 
The point $x_\ast$, which coincides with the stationary point of the equation for the
quantity $\Delta= Y-Y_0$, is where the maximum departure of the WIMPs abundance $Y$ from the thermal value $Y_0$ is reached.
For each  mass $m_\chi$ and total annihilation cross section $\langle \sigma_\text{ann}v_\text{r}\rangle$, 
the temperature $x_\ast$ and the actual WIMPs abundance $Y(x_\ast)$ are exactly known. This value
provides the true  initial condition for the usual differential equation that have to be integrated 
in the interval $x\ge x_\ast$. 
The matching of the two abundances at $x_\ast$ is continuous and differentiable. 
The dependence of the present relic abundance on the abundance at an intermediate temperature is an exact result.
The exact theory suggests a new analytical approximation that furnishes  the relic abundance accurate at the level of $1\%-2\%$ in the case of $S$-wave and $P$-wave scattering cross sections. We conclude the paper studying the evolution of the WIMPs chemical potential and the entropy production using methods of non equilibrium thermodynamics.

\end{abstract}

\pacs{95.35.+d,98.80.-k,04.20.Cv}
	

\maketitle

\section{Introduction and motivations}
Freeze out~\cite{ZOP,LW} refers to the idea that stable particles, once in thermal and chemical equilibrium in 
the hot and dense early Universe, 
left a relic abundance because at a certain stage of the evolution, the expansion and cooling made their density so 
small that the annihilation reaction rates became frozen.
Earlier reviews on
the connection between cosmology and particle physics are found in  Refs.~\cite{Steigman79,DolgovZeldovich}. 

Today we know that  most of the mass of the  Universe is  constituted by some form of non luminous 
dark matter as evidenced by the  determination  
of the cosmological parameters from the data provided by the WMAP and Planck satellites~\cite{experiments}.
A massive, neutral, stable particle that interacts with the standard model particles
with a strength similar to that of weak interaction (weakly interacting massive particles or WIMPs), 
is the paradigmatic relic and could explain the properties of dark matter.

Using natural units where $\hslash=c=k_B=1$, the well known~\cite{ZOP,LW} rate equation for the number 
density $n$ of relics is 
\begin{equation}
\frac{1}{a^3}\frac{d(n a^3 )}{dt}=\frac{dn}{dt}+3Hn=\langle \sigma_{\text{ann}} v_\text{r} \rangle  (n^2_{0}-n^2),
\label{ZOPequation}
\end{equation}
where $a$ is the Friedmann-Robertson-Walker scale factor, $a^3$ the comoving volume,
$H=({1}/{a}) {da}/{dt}$ the Hubble parameter,
$\langle \sigma_{\text{ann}} v_\text{r} \rangle$ the total 
thermally averaged annihilation unitary rate, and $n_0$ the number density at zero chemical potential
determined by the equilibrium statistics obeyed by WIMPs. 

It is useful to write 
Eq.~(\ref{ZOPequation}) in a different form~\cite{KO,ScherrerTurner,GG,EG}.
If the expansion proceeds adiabatically, the entropy per comoving volume $S=s a^3$ is conserved, 
with entropy density $s=(2\pi^2/45)g_s T^3$.
In the radiation dominated epoch $H=\sqrt{(8/3) \pi G \rho}$, with $M_P=1/\sqrt{G}=1.22\times 10^{19}$ GeV the Planck mass
and $\rho=(\pi^2 /30) g_\rho T^4$ the energy density. 
Using the abundance $Y=N/S=n/s$ as a function of $x=m_\chi /T$ instead of the time, 
with the change of variable $d/dt =H x d/dx$, the equation becomes~\cite{KO,ScherrerTurner,GG,EG}
\begin{flalign}
\frac{dY}{dx}&= 
\frac{C}{x^2} {\langle \sigma_{\text{ann}} v_\text{r}\rangle}(Y^2_{0}-Y^2),
\label{Yequation}
\\
C&=\sqrt{\frac{\pi}{45}} M_P m_\chi \sqrt{g_*},
\nonumber
\end{flalign}
where $\sqrt{g_*} ={g_s}/ {\sqrt{g_\rho}} (1+{T}/{3}\,  d(\ln g_s )/dT)$
accounts for the temperature dependence of the relativistic degrees of freedom 
$g_\rho$ and $g_s$~\cite{OliveScrammSteig,SWO,GG,Steigman2012}.

The present relic abundance is given by $Y(x_0)$,
which is the constant asymptotic value of the solution of Eq.~(\ref{Yequation}),
with $T_0=2.725$ K, the temperature of the microwave background today.
The dark matter relic mass density is $\rho_{\text{DM}}=m_{\chi} Y(x_0) s_0$
with $s_0 =2890$ cm$^{-3}$ the entropy density today. 
The ratio over the critical density $\rho_c =3H^2_0 /(8\pi G)$ is 
$\Omega_{\text{DM}} h^2=2.75\times 10^{8} \left({m_\chi}/{\text{GeV}}\right) Y(x_0)\simeq 0.11$,
where $h=H_0 / (100\, \text{km}\, \text{s}^{-1}\, \text{Mpc}^{-1})$ is the reduced Hubble constant. 
The numerical value of the ratio is the present experimental central value. 

To a very good approximation, quantum statistical effects can be neglected~\cite{GG}, 
thus the thermal number density for non degenerate 
relativistic gas is given by the Maxwell-Boltzmann-Juttner
statistics~\cite{KolbWolfram} with zero chemical potential  $\mu_i = 0$,
\begin{flalign}
n_{0}= \frac{g}{2\pi^2} T^3  x^2  K_2 (x),\;\;\;\;
Y_{0}=\frac{45}{4 \pi^4}\frac{g}{g_s}
x^2 K_2 (x),
\label{n_0}
\end{flalign}
Here $g$ are the spin degrees of freedom  of the particle
and  $K_i$ are modified Bessel functions of the second kind. 
In the relativistic limit $x=0$ become $n^{m=0}_{0}= ({g}/{\pi^2}) T^3$, $Y^{m=0}_0=45/(2\pi^4) {g_\chi}/{g_s}$.

An approximate estimation of the relic density can be obtained with
the approximate analytical solution of Eq.~(\ref{ZOPequation}), which is known as \textit{freeze out 
approximation}, already formulated  by Zeldovich, Okun and Pikelner  in Ref.~\cite{ZOP}. 
The evolution can be divided into two stages:
in the first stage, at earlier times during the expansion, WIMPs go slowly out of chemical equilibrium, thus 
$n \sim n_0$. 
The relative variation of the density because of the expansion has characteristic time  
$t^{-1}_{1} =1/(n_0 a^3) d(n_{0} a^3)/dt$, while the characteristic 
time of interactions in equilibrium is $\tau^{-1} = 
\langle \sigma_{{\text{ann}}} v_{\text{r}} \rangle n_{0}$.
The instant $t_f$ of "freezing"~\cite{LW}, or "quenching"~\cite{ZOP}, separating the two stages 
is estimated by requiring $t^{-1}_1=\tau^{-1}$.
In modern notation, we can write
\begin{flalign}
-\frac{1}{Y_0 } \frac{dY_{0}}{dx}&\simeq   \frac{C}{x^2}\langle\sigma_{{\text{ann}}}{\text{v}}_{\text{r}} \rangle Y_{0}
\;\;\text{at $x=x_f$}.
\label{Zeld_criterion_1}
\end{flalign}
We have inserted a minus sign because the derivative of $Y_0$ is always negative.
In the second stage, at later times, the creation term
proportional to $Y^2_0$ in (\ref{Yequation}) can be neglected and the equation 
$dY/dx=-(C/x^2) {\langle \sigma_{\text{ann}} v_\text{r}\rangle}Y^2$ is easily integrated
with the initial condition
\begin{flalign}
Y(x_f)&\simeq 2 Y_0(x_f).
\label{Zeld_criterion_2}
\end{flalign}
Equation~(\ref{Zeld_criterion_2}) is used in 
Refs.~\cite{ZOP,DolgovZeldovich,VDZ} as the approximate 
initial condition for the integration. The origin of the factor 2 is not explained.
In this respect, we note that  the freeze out condition used by Lee and Weinberg~\cite{LW} and successive 
earlier works~\cite{Hut,Wolfram,Goldberg,Krauss,Ellis,SZ}
is equivalent to Eq.~(\ref{Zeld_criterion_1}), but in the initial 
condition for the integration there is not the factor 2 as in (\ref{Zeld_criterion_2}).
We will discuss this point in Section~\ref{Sec_relic_approx}.

The criterion (\ref{Zeld_criterion_1}) takes a more familiar form in the nonrelativistic limit.
From Eq.~(\ref{n_0}) it is easy to see that 
\begin{flalign}
-\frac{1}{Y_0}\frac{d Y_{0}}{dx}=
\frac{K_1(x)}{K_2(x)}
\overset{\text{\tiny $x \gg 1$}}{\longrightarrow}1.
\label{log_der}
\end{flalign}
Using $C/x^2 =s/(Hx)$ and $\Gamma=\langle\sigma_{{\text{ann}}}{\text{v}}_{\text{r}} \rangle Y_{0} s$, 
Eq.~(\ref{Zeld_criterion_1}) becomes
\begin{flalign}
\frac{\Gamma}{Hx}\sim 1\;\;\text{at $x=x_f$},
\label{popular_criterion}
\end{flalign}
which is the more precise version (see the Erratum of~\cite{ScherrerTurner}) of the popular criterion "the 
annihilation rate per particle $\Gamma$ equals the expansion rate $H$".

The exact solution of the non-linear Riccati 
equation (\ref{Yequation}), on the other hand, in general can only be obtained with numerical methods. 
In order to do that, an initial condition is necessary.
The differential equation is first order, thus its solution is determined completely by the initial condition.
The true initial condition is $Y(0)=Y_0 (0)$, which  anyway is only an asymptotic condition because the 
equation has a singularity in $x=0$. 
The numerical integration hence is done by setting $Y=Y_0$ at some small $x$, for example $x=2$~\cite{EG},
where the solution is expected to be very close to $Y_0$.  
It must be stressed that this is a false initial condition 
because $Y>Y_0$ at any $x>0$. The difference $Y-Y_0$ at small $x$ 
is anyway so small that the numerical result is correct for all practical purposes.

If the differential equation gives the actual abundance in all the range $(0,\infty)$, the freeze 
out temperature $x_f$ does not exist, is an artificial arbitrary point introduced by the freeze out 
approximation~\cite{BenderSarkar}.
Also the "exact" solution is in reality an approximation because uses a false initial condition.

The problem of the initial condition is important for the relic abundance calculation
because it determines the precision with which the asymptotic value is obtained.
For example, public codes use very 
different approaches. In \textsf{DarkSUSY}~\cite{darksusy} the numerical integration is started 
with $Y(x=2)=Y_0(x=2)$ with a step-size 
changing adaptive implicit method to avoid  stiffness problems. \textsf{MicrOMEGAs}~\cite{micromegas} 
starts  at the point $x_{f_1}$ where 
$Y-Y_0=0.1 Y_0$ performing a Runge-Kutta integration up to the point $x_{f_2}$ where
$Y(x_{f_2})> 10 Y_0 (x_{f_2})$;  then the analytical approximation neglecting $Y_0$ is employed. 
\textsf{SuperIso}~\cite{superiso} uses the freeze out 
approximation as given of Ref.~\cite{GG}. In \textsf{MadDM}~\cite{maddm} 
the asymptotic value is calculated integrating the equation from $x=30$ up to a large vale, $x=1000$. 
The process is repeated backward 
varying the initial $x$ until $(Y_\infty (x_i)-Y_\infty (x_{i-1}))/Y_\infty (x_i)$
is less than the required precision. 
It is clear that these codes, leaving aside the uncertainties intrinsic in the numerical methods, 
for a given WIMP mass and total 
annihilation cross section inevitably will differ in the value
of the relic abundance, even if within a few percent in normal situations.

The problem that we pose and  answer in this paper hence is the following.
Suppose that for each WIMP mass and total annihilation cross section the actual abundance is exactly known 
in one intermediate point $x_\ast$. Then we can  use it as true initial condition for the differential equation,
and  the asymptotic value at large $x$ calculated in this way would be the true  value.
But a conceptual problem arises:
it would  tell us that the relic abundance really depends on an intermediate temperature, in a way similar to
the freeze out approximation, and not on the asymptotic value at $x=0$ as it should be if the first order 
differential equation described the evolution all over the range. In other words, in that case 
the usual differential equation would have a boundary value at $x_\ast$, the evolution before $x_\ast$
would be given by something else.

In the next section we show that such a point exists.

\section{A special temperature}

In present literature it is typically adopted the freeze out approximation  in the form given by 
Scherrer and Turner~\cite{ScherrerTurner}, which is similar to the earlier treatment of 
Steigman~\cite{Steigman79,Steigman2012}. 
Calling $\lambda(x)={C}\langle \sigma_{\text{ann}}v_\text{r}   \rangle/{x^2}$,
the method  consists in studying the differential equation for the difference  
$\Delta=Y-Y_{0}$, the "distance" from equilibrium,
\begin{flalign}
\frac{d\Delta}{dx}=-\frac{dY_0}{dx}-\lambda(x)\Delta (2Y_0 +\Delta).
\label{Delta_equation}
\end{flalign}
The  matching point of the two approximate solutions in 
the two regimes is found with the ansatz
$\Delta (x_f)=c Y_{0} (x_f)$, 
being $c$ an arbitrary constant numerical factor 
$\mathcal{O}(1)$ determined by fitting the 
numerical solution of Eq.~(\ref{Yequation}) for various 
masses and cross sections.
Values for $c$ such as $\sqrt{2}-1$~\cite{ScherrerTurner,GS}, 
$1/2$~\cite{Steigman79,mazumdar}, $(\sqrt{5}-1)/2$~\cite{Steigman2012},
$1.5$~\cite{GriestSeckel86,GG},
has been found to provide the relic abundance accurate at the level of a few percent.
The value of $x_f$
is determined neglecting $d\Delta /dx$ in (\ref{Delta_equation}) and substituting $\Delta (x_f)=c Y_{0} (x_f)$.
It depends logarithmically on $c$. 
The equation is numerically stiff, thus is not surprising that different 
authors find different values for numerical coefficients $c$.

In reality Eq.~(\ref{Delta_equation}) tells something much more precise.
Equation (\ref{Delta_equation}) posses non trivial 
stationary points where ${d\Delta}/{dx}=0$.
The difference $Y-Y_0$ is initially zero, then grows (always positive) and tends to $Y$ asymptotically
because $Y_0$ is rapidly decreasing. Hence the extremal point is a maximum. We indicate the stationary 
point with $x_\ast$ to distinguish it from the arbitrary point $x_f$ of the freeze out approximation. 
At $x=x_\ast$ Eq.~(\ref{Delta_equation}) is a quadratic algebraic equation with constant coefficients,
\begin{flalign}
\Delta^2 (x_\ast) + 2 Y_0 (x_\ast) \Delta(x_\ast) +\frac{1}{\lambda(x_\ast)} \frac{dY_0}{dx}\Big|_{x_\ast}=0.
\end{flalign}
Disregarding the non-physical negative solution, and indicating with a subscript $\ast$ all the quantities 
evaluated at $x_\ast$, the physical solution is
\begin{flalign}
\Delta_\ast &=-Y_{0,\ast} +\sqrt{Y^2_{0,\ast} -\frac{1}{\lambda_\ast} \frac{dY_0}{dx}\Big|_\ast},
\label{Delta_at_xf}
\end{flalign}
and the abundance at $x_\ast$ then takes the form
\begin{flalign}
Y_\ast & =\sqrt{Y^2_{0,\ast}-\frac{1}{\lambda_\ast} \frac{dY_0}{dx}\Big|_\ast}.
\label{Y_at_xf} 
\end{flalign}
If we now define 
\begin{flalign}
c_\ast=\sqrt{1- \frac{1}{\lambda_\ast Y_{0,\ast}}
\frac{1}{Y_{0,\ast}} \frac{dY_{0}}{dx}\Big|_\ast} -1,
\label{delta_xf} 
\end{flalign}
then we have
\begin{flalign}
\Delta_\ast& = c_\ast Y_{0,\ast},\\
Y_\ast &= (1+c_\ast) Y_{0,\ast}.
\label{ansatzexact}
\end{flalign}
At $x=x_\ast$, then, the ansatz 
is an exact result with $c=c_\ast$. 
Now $c$ is not an arbitrary numerical value, but it is determined in each
case by the WIMP mass, the total annihilation cross section and $Y_0 (x)$.
\footnote{Applying the criterion (\ref{popular_criterion}) in the square root of  Eq.~(\ref{delta_xf}) we 
obtain $c=\sqrt{2}-1$, that is one of values found by numerical integration and fitting.
This also explains why using non relativistic $n_0$ one finds values around $0.5$ for the 
numerical constant $c$.}

At $x_\ast$ the actual abundance $Y$ and $c$ have a well defined analytical form as seen in Eqs.~(\ref{Y_at_xf}) and (\ref{delta_xf}). We call the functions defined by those formulas
 $Y_1(x)$ and $\delta(x)$, 
\begin{flalign}
Y_1(x)&=\sqrt{Y^2_0 - 
\frac{x^2}{C \langle \sigma_{\text{ann}} \text{v$_\text{r}$}\rangle  }
\frac{dY_{0}}{dx}}= (1+\delta(x))Y_0 (x),
\label{Y1}
\\
\delta (x)&=\sqrt{1- 
\frac{x^2}{C \langle \sigma_{\text{ann}} \text{v$_\text{r}$}\rangle Y_{0}  }
\frac{1}{Y_0} \frac{dY_{0}}{dx}} -1.
\label{delta}
\end{flalign}
The point $x_\ast$ where $d\Delta/dx=0$ is evidently given by
$dY_1/dx|_{x=x_\ast} =dY_0 /dx|_{x=x_\ast}$, or, using Eqs.~(\ref{Y1}), (\ref{delta}),
\begin{flalign}
-\frac{1}{Y_0(x)}\frac{dY_0(x)}{d x}=
\frac{1}{\delta(x)} \frac{\delta(x)}{dx}\,\, \text{at $x=x_\ast$}.
\label{freezeoutcondition}
\end{flalign}
The root of Eq.~(\ref{freezeoutcondition}) gives $x_\ast$.
Using Eq.~(\ref{Y_at_xf}) also the actual abundance at the point $x_\ast$ is known.
The differential equation (\ref{Yequation}) can be solved in the interval $x\geq x_\ast$
with the true initial value $Y(x_\ast)=Y_1(x_\ast)$. 
We call $Y_2 (x)$ the actual abundance $Y(x)$ at $x\ge x_\ast$,
\begin{flalign}
\frac{dY_2}{dx}&= 
\frac{C}{x^2} {\langle \sigma_{\text{ann}} \text{v$_\text{r}$}\rangle}(Y^2_{0}-Y^2_2),
\label{Y2eq}
\\
Y_2 (x_\ast)&=Y_1 (x_\ast).
\label{initialcondition}
\end{flalign}

As far as the calculation of the relic density is concerned, the problem where to start the integration
is solved: for each mass and total annihilation cross section, Eqs.~(\ref{freezeoutcondition}) and (\ref{Y1})
furnish the true initial condition, what happens before $x_\ast$ does not matter. There is no reason to start
the integration before (or after) $x_\ast$ with a false initial condition.
We remark that there is no approximation in this procedure.

But if the evolution before $x_\ast$ does not affect the abundance after $x_\ast$, it means that 
the differential evolution in temperature really starts at $x_\ast$, while
at higher temperatures it should be fixed by some constraint.

Let us look at the properties of the function $Y_1(x)$.
The function $Y_1(x)$ is not solution of the differential equation, as it is easily seen by
substitution, and  it has the correct behavior to describe the evolution
before $x_\ast$. In fact, it tends to $Y_0$ for $x \to 0$ asymptotically.
At small $x$ we may expand the square root 
and  find 
\begin{flalign}
Y_1 \sim Y_0 - \frac{1}{2} \frac{x^2}{C \langle \sigma_{\text{ann}} \text{v$_\text{r}$}\rangle 
Y_0}\frac{dY_0}{dx},
\end{flalign}
which is the usual freeze out approximation~\cite{ScherrerTurner,BBF85}.
Furthermore, and more important, given that $dY_2/dx=dY_1/dx=dY_0/dx$ at $x=x_\ast$,
the piecewise function 
\begin{flalign}
Y(x)=\left\{  \begin{array}{lr}
Y_1(x),\; x\leq x_\ast
\\   Y_2(x),\;x\geq x_\ast,
 \end{array}
   \right.
   \label{piecewiseY}
\end{flalign}
is continuous and differentiable in $x_\ast$ as it must 
be for the exact evolution. In the freeze out approximation, on the contrary, 
$x_f$ is an elbow point.

Then the natural questions:
is $Y_1(x)$ the exact actual abundance at $x\le x_\ast$? 
Where does it come from?

\section{Exact theory}

Usually  Eq.~(\ref{ZOPequation}) is derived from the Boltzmann equation~\cite{GG,BGS,BBF85,books}.
The integrated  Boltzmann equation~(\ref{ZOPequation}) has  the form  
of a kinetic chemical equation, which suggests to approach the  problem from the point of view 
rate equations that describe non equilibrium chemical reactions. This thermodynamic approach allows to arrive 
at the results and answer the previous questions more easily. 

\subsection{Generalities on rate equations}

Let $\nu_1 X_1 + \nu_2 X_2 \rightleftarrows \nu_3 X_3 +\nu_4 X_4$ be a 
an elementary stoichiometric  $2\to 2$ reaction between species $X_i$ with stoichiometric coefficients $\nu_i$
that can proceed at the same time in both directions.
We follow the convention that the stoichiometric coefficients $\nu_i$ are  negative for reactants $X_{1,2}$
and positive for products $X_{3,4}$. 
It is convenient to consider the concentration given by the number density
$n_i={N_i}/{ V}$, with $N_i$ the numbers of particles 
instead of the numbers of moles in a volume $V$. 

The forward (left to right) and backward (right to left) rates are
\begin{flalign}
\mathcal{R}_f=k_f \;n^{|\nu_1|}_1 n^{|\nu_2|}_2,\;\;\;\;
\mathcal{R}_b=k_b \;  n^{|\nu_3|}_3 n^{|\nu_4|}_4,
\label{for_back_rates}
\end{flalign}
where $k_f$, $k_b$ are the related temperature dependent rate constants.
In chemical equilibrium, for the detailed balance principle, the rates 
are equal, $\mathcal{R}_f=\mathcal{R}_b$, and the concentrations stay constant at their equilibrium value  
that defines the equilibrium constant $K_{\text{eq}}(T)={k_f}/{k_b}$.

If concentration changes because both the number of particles, due to reactions, and the volume, due to expansion,  
change in time, then the rate equation for the specie $X_i$ is given by
\begin{flalign}
\frac{1}{\nu_i} \frac{1}{V}\frac{d (n_i V)}{dt}=\mathcal{R}_f-\mathcal{R}_b .
\label{rate_equation_chemical}
\end{flalign}

Consider  $\chi$, a self-conjugated neutral 
particle that annihilates into particle-antiparticles pairs of the thermal bath constituted by $N_b$ species $\psi_i$, 
$i=1,...,N_b$ of the standard model. The $\psi$ particles are all the particles that interact with the WIMP in a given model,
including massless particles like the photons. In this case clearly $\psi=\bar{\psi}$. 
We further assume that the only reactions that change the number of WIMPs is  pair annihilation and 
pair-creation, 
\begin{flalign}
2\,\chi \rightleftarrows \psi_i  {\bar{\psi}_i}.
\nonumber
\end{flalign}
In the case of $2\to 2$ collisions in a gas of elementary particles, 
the temperature dependent rate is given by
\begin{flalign}
\langle \mathcal{R}\rangle=\frac{n_1  n_2 }{1+\delta_{12}}
 \int dv_\text{r} P(v_\text{r} )
\sigma v_\text{r}
=\frac{n_1  n_2 }{1+\delta_{12}}
{\langle \sigma v_\text{r}\rangle}.
\nonumber
\end{flalign}
The factor $1/(1+\delta_{12})$ accounts for the fact that 
if 1 and  2 are the same species and can react, then  each reaction is counted twice.
Here $v_\text{r}$
is the relative velocity of the colliding particles and  $P(v_\text{r} )$ its probability 
distribution. In the case of relativistic particles obeying the Maxwell-Boltzmann-Juttner statistics, 
the probability distribution of the relativistic relative velocity was discussed in Ref.~\cite{Cannoni}.

\subsection{Standard assumption for thermal relics}

In the early Universe temperature was so high that
all fields behaved as thermal radiation with zero chemical potential, i. e. the
number density of particles depends only on the temperature as given by Eq.~(\ref{n_0}).
The reactions $2\,\chi \rightleftarrows \psi_i  {\bar{\psi}_i}$ at the initial time
were in chemical equilibrium, 
\begin{flalign}
&\mathcal{R}^{\text{eq}}_{f,i} =\mathcal{R}^{\text{eq}}_{b,i}\Longrightarrow
\frac{\langle \sigma _{\chi \chi} 
v_\text{r} \rangle_i}{2} n^2_{0,\chi}
=\langle \sigma_{\psi_i  {\bar{\psi}_i}} v_\text{r} \rangle n^2_{0,\psi_i}.
\label{assumption_1}
\end{flalign}
While the number density of the bath particles is much larger than that of the WIMPs and
maintains its thermal form during the freeze out process
in reason of the fast interactions between them,
\begin{flalign}
n_{\psi_i}=n_{\bar{\psi}_i}=n_{0,\psi_i},
\label{assumption_2}
\end{flalign}
WIMPs have to go out of thermal equilibrium, $n_\chi \neq n_{0,\chi}$. Altogether also 
chemical equilibrium would be maintained, with densities changing only because of the 
expansion/temperature variation and there would be no relic abundance.

Note that assumption (\ref{assumption_2}), equilibrium statistics with  chemical potential maintained to zero,
means that the bath particles behave like thermal photons of the
black body radiation, i.e. their density is completely fixed
by the temperature and does not depend on the amount of material of the cavity, 
which in our  case is  the amount of WIMPs in the comoving volume. 
As a consequence, in the process of thermal WIMP production and decoupling
the total  number of particles is not conserved.

\subsection{Rate equation for WIMPs}

Using the above formalism with $V=a^3$, 
and the notation $\langle \sigma_{\chi\chi} v_\text{r} \rangle_i$
and $\langle \sigma_{\psi_i  {\bar{\psi}_i} } v_\text{r} \rangle$ for the annihilation 
and creation rates, respectively, we can write the rate
equation for WIMPs,
\begin{flalign}
-\frac{1}{2}\frac{1}{a^3}
\frac{d(n_{\chi} a^3)}{dt} =
\langle \sigma_{\chi\chi} v_\text{r} \rangle_i \frac{n^2_{\chi}}{2}
-\langle \sigma_{\psi_i  {\bar{\psi}_i} } v_\text{r} \rangle n_{\psi_i} n_{\bar{\psi}_i}.
\label{rate_eq_wimps}
\end{flalign}
The assumption (\ref{assumption_2}), together with Eqs.~(\ref{assumption_1}), 
implies that the  creation rates are determined  by thermal densities:
\begin{flalign}
\langle \sigma_{\psi_i  {\bar{\psi}_i} } v_\text{r} \rangle n_{\psi_i} n_{\bar{\psi}_i}
\overset{\text{\tiny Eq.(\ref{assumption_2})}}{=}
\langle \sigma_{\psi_i  {\bar{\psi}_i}} v_\text{r} \rangle n^2_{0,\psi_i}
\overset{\text{\tiny Eq.(\ref{assumption_1})}}{=}
\frac{\langle \sigma _{\chi \chi} 
v_\text{r} \rangle_i}{2} n^2_{0,\chi},
\label{substitution_rate} 
\end{flalign}
Doing these substitutions in Eqs.~(\ref{rate_eq_wimps}) 
\begin{flalign}
-\frac{1}{2}\frac{1}{a^3}
\frac{d(n_{\chi} a^3)}{dt} =
\langle \sigma_{\chi\chi} v_\text{r} \rangle_i \frac{n^2_{\chi}}{2}
-\frac{\langle \sigma _{\chi \chi} 
v_\text{r} \rangle_i}{2} n^2_{0,\chi},
\label{rate_eq_wimps_1}
\end{flalign}
We can now define the total annihilation cross section
\begin{flalign}
\langle \sigma_{\text{ann}} v_\text{r}\rangle=\sum^{N_b}_{i=1} \langle \sigma _{\chi \chi} 
v_\text{r} \rangle_i,
\label{tota_sigamann}
\end{flalign}
and sum over the channels. Eq.~(\ref{rate_eq_wimps_1}) finally reads  
\begin{flalign}
\frac{1}{a^3}\frac{d(n_\chi a^3)}{dt}=\langle \sigma_{\text{ann}} v_\text{r} \rangle
(n^2_{0,\chi}-n^2_\chi),
\label{eq2}
\end{flalign}
which  is the usual differential equation for WIMPs\footnote{
Note the cancellation between the stoichiometric and the 
statistical factors~\cite{Ellis}.
Had we started with a particle-antiparticle WIMP system with equal abundances,
there would be another equation for $\bar{\chi}$ that is equal to the one for $\chi$, 
with stoichiometric and rate statistical factors equals to one. 
The doubling of all the terms when summing the two equations cancels, hence  Eqs.~(\ref{eq1}) and (\ref{eq2})  
are obtained again.}  derived under the usual assumptions.
\subsection{Equation for bath particles}

Let us pass now to the equations for the bath particles that read
\begin{flalign}
\frac{1}{a^3}\frac{d(n_{\psi_i} a^3)}{dt} =
\langle \sigma_{\chi\chi} v_\text{r} \rangle_i \frac{n^2_{\chi}}{2}
-\langle \sigma_{\psi_i  {\bar{\psi}_i}} v_\text{r} \rangle n_{\psi_i} n_{\bar{\psi}_i}.
\label{rate_eq_bath1}
\end{flalign}
The rate equation for $\bar{\psi}_i$ coincides with Eq.~(\ref{rate_eq_bath1}) setting $\bar{\psi}_i$ in the
time derivative, thus we do not write it explicitly.
Given that the equations for $n_{\psi_i}$ and $n_{\bar{\psi}_i}$ are similar
we sum Eq.~(\ref{rate_eq_bath1}) over the $ N_b$ channels and multiply by 2. Employing the assumption
(\ref{assumption_2}), $n_{\psi_i}=n_{\bar{\psi}_i}=n_{0,\psi_i}$, we obtain
\begin{flalign}
\sum_{i=1}^{N_b} [2\frac{d(n_{0,\psi_i} a^3)}{dt} =
2 a^3 \langle \sigma_{\chi\chi} \text{v$_\text{r}$} \rangle_i \frac{n^2_{\chi}}{2}
-2 a^3 
\langle \sigma_{\psi_i  {\bar{\psi}_i}} \text{v$_\text{r}$} \rangle n^2_{0,\psi_i} ].
\label{eq_bath_before_simpl_1}
\end{flalign}
Using now (\ref{substitution_rate}) and (\ref{tota_sigamann}) as we did for the WIMPs, 
Eq.~(\ref{eq_bath_before_simpl_1}) results in
\begin{flalign}
\frac{2}{a^3}\sum_{i=1}^{N_b} \frac{d(n_{0,\psi_i} a^3)}{dt} =-
 \langle \sigma_{\text{ann}} v_\text{r} \rangle
 (n^2_{0,\chi}-n^2_\chi).
\label{eq_bath_before_simpl}
\end{flalign}
Let us pause and look at this equation.

Already at this stage it is clear that this is an algebraic equation, not a differential equation. In fact, 
the only unknown is the actual density $n_\chi$ of the WIMPs, which appears only 
on the right-hand side. The derivative on the left-hand 
side is the sum of the time/temperature derivative
$n_{0,\psi_i}$  for each of the $N_b$
species. The left hand side of (\ref{eq_bath_before_simpl}) 
is hence a known function as much as $n_{0,\chi}$. 

If the bath particles had a chemical potential, hence their density $n_{\psi,i}$ would evolve in time 
as dictated by the rate equation in a similar way to $n_\chi$. 
Conservation of the total number of particles would hold 
and we would have $2\sum_{i=1}^{N_b} {d(n_{\psi ,i} a^3)}/{dt}
=-{d(n_{\chi} a^3)}/{dt}$. Equation (\ref{rate_eq_bath1}) would give the same Eq.~(\ref{rate_eq_wimps}).
We would have only one equation, but the creation term would be 
$\sum\langle \sigma_{\psi_i  {\bar{\psi}_i}} v_\text{r} \rangle n_{\psi_i} n_{\bar{\psi}_i}$
and could not be expressed in terms of the annihilation cross sections and of the thermal equilibrium
density of the WIMPs.

Anyway, by hypothesis, the bath particles are locked to the thermal state with zero chemical potential, i.e.
the number density of the bath particles depends only on the temperature
and is not affected by the reactions with the WIMPs.
The total number of particles is not conserved: the time/temperature derivative $2\sum_{i=1}^{N_b} 
d(n_{0,\psi_i} a^3)/dt$ cannot be equal to $-d(n_{\chi} a^3)/{dt}$. 

Can we further simplify the left-hand side of Eq.~(\ref{eq_bath_before_simpl})? 
Note that the expansion changes the density $n_0$ 
but the product $n_0 a^3$ is not affected. 
At the same time 
the thermal density $n_0$ depends only on temperature and is not changed by the reactions.
The quantity
\begin{flalign}
(2\sum_{i=1}^{N_b} n_{0,\psi_i} +n_{0,\chi})a^3,
\label{crucial_0}
\end{flalign}
then remains constant during the freeze out process. Hence 
\begin{flalign}
{2}\sum_{i=1}^{N_b} \frac{d(n_{0,\psi_i} a^3)}{dt} =-\frac{d(n_{0,\chi} a^3)}{dt},
\label{crucial}
\end{flalign}
which substituted in Eq.~(\ref{eq_bath_before_simpl}) finally gives
\begin{flalign}
\frac{1}{a^3}\frac{d(n_{0,\chi} a^3)}{dt}=
\langle \sigma_{\text{ann}} v_\text{r} \rangle
(n^2_{0,\chi}-n^2_\chi).
\label{eq1}
\end{flalign}
In terms of the abundance $Y=n/s$ and dropping the subscript $\chi$, 
Eqs. (\ref{eq1}) become
\begin{flalign}
\frac{dY_0}{dx}= 
\frac{C}{x^2} {\langle \sigma_{\text{ann}} \text{v$_\text{r}$}\rangle}(Y^2_{0}-Y^2),
\label{Y1eq}
\end{flalign}
which solved for $Y$ directly gives the function $Y_1$ of Eq.~(\ref{Y1}).

Given that the algebraic equation (\ref{Y1eq}) and the differential equation (\ref{Yequation})
must hold at the same time and the right hand side is the same, besides the trivial solution $Y_1(x) =Y_2 
(x)=Y_0(x)=Y^{m=0}_0$, 
the only other possibility is that the function $Y_1$ gives the actual abundance in the interval $(0, x_\ast]$, 
while the function $Y_2$ gives the actual abundance in the interval $[x_\ast,\infty)$. The solutions can coincide only at 
point $x_\ast$
and the matching is continuous and differentiable as we have already seen.   

The functional form of the abundance discussed in Section II, Eq.~(\ref{Y_at_xf}), hence, 
is the true non equilibrium WIMPs abundance at $x \leq x_\ast$.

\subsection{Numerical Example}

In order to show how this works in practice,  we take $m_\chi =100$ GeV, 
$g_\chi =2$, $\langle\sigma_\text{ann} v_\text{r}\rangle=10^{-10}$ GeV$^{-2}$ and $g_s =g_\rho=g=100$, 
$\sqrt{g_*}=\sqrt{g}$, neglecting the temperature dependence of the degrees of freedom.

\begin{figure}[t!]
\includegraphics*[scale=0.58]{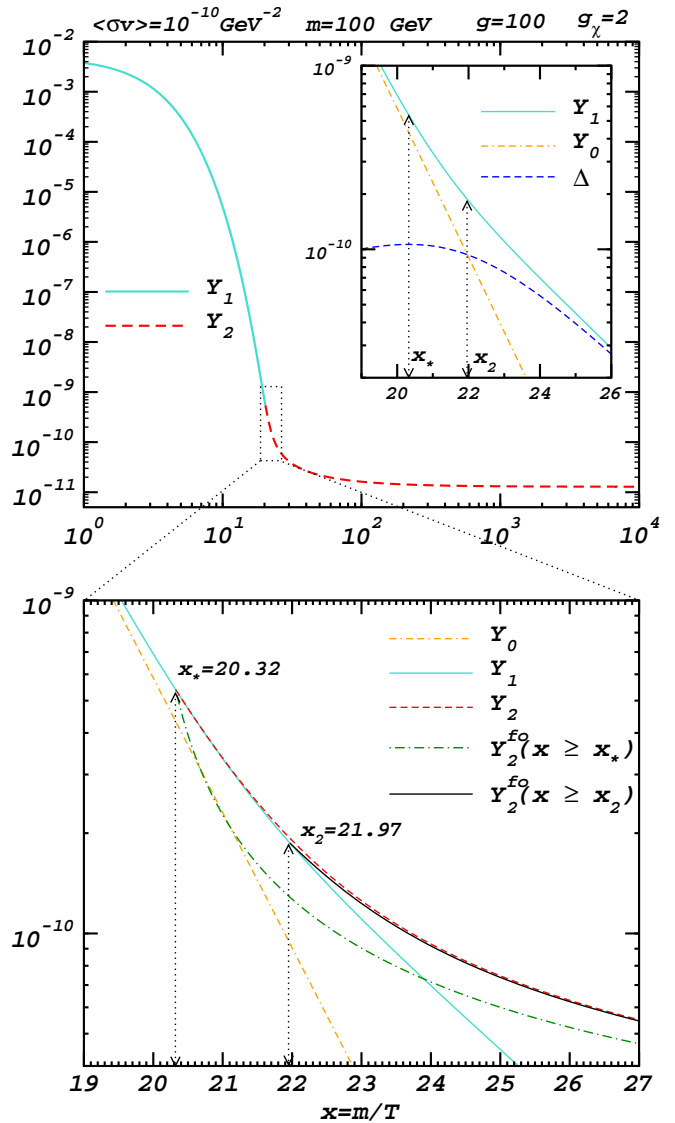}
\caption{\textit{Top panel:} 
The piecewise function build with $Y_1( x\le x_\ast)$, 
Eq.~(\ref{Y1}), and $Y_2( x\ge x_\ast)$, the numerical 
solution of (\ref{Y2eq})-(\ref{initialcondition})
with $x_\ast$ given by Eq.~(\ref{freezeoutcondition}).
The \textit{inlay panel} evidences that $x_\ast$ 
corresponds to the position of the maximum of 
$\Delta$, and $x_2$ is the intersection of $\Delta$ with $Y_0$. 
\textit{Bottom panel:} The freeze out zone evidenced with a box in the top panel.
It is also shown $Y_0$ and the analytical approximation
(\ref{Y2_app}) starting both at $x_\ast$ 
and at $x_2$. 
}
\label{Fig1}
\end{figure}

From Eq.~(\ref{freezeoutcondition}) we find that the freeze-out temperature is $x_\ast =20.32$. 
The inlay in the top panel of Figure~\ref{Fig1} shows the behavior of $\Delta$ and the position of its maximum.
In the top panel of Figure~\ref{Fig1} we show the piecewise function built with $Y_1 (x\le x_\ast)$, 
Eq.~(\ref{Y1}), solid turquoise curve, 
and $Y_2 ( x\ge x_\ast)$, dashed-red curve, 
obtained by  numerical integration of Eq.~(\ref{Y2eq}) with the initial condition (\ref{initialcondition}).
The standard picture is obtained. 

The actual abundance $Y_1$ in the early stage at small $x$ closely tracks the equilibrium 
function $Y_0$ because the term $-x^2/(C \langle \sigma_{\text{ann}} v_\text{r}\rangle )dY_0/dx$  
is very small. Only with $x$ approaching $x_\ast$ it becomes relevant;
numerically we find that $\delta(x_\ast)=0.24$ but, for example, $\delta(2)\sim 10^{-10}$. 
The smallness of $\delta(2)$ explains why if one integrates the differential equation with the false
initial condition $Y(2)=Y_0 (2)$, once obviated the stiffness problems, one obtains a curve that is
practically indistinguishable from that in Fig.~(\ref{Fig1}). 

The numerical advantages of the exact theory should be clear: integrating the differential equation 
from $x_\ast$, the solution varies smoothly less that two orders of magnitudes, instead of eight,
before reaching the asymptotic 
constant value: there is no stiffness problem. The asymptotic value obtained in this way is the true value.

\subsection{Discussion}

If one writes down the Boltzmann equations for the bath particles and simplify
them under the same assumptions used in deriving the equation for the WIMPs, one would arrive 
at the same results. 
The crucial point is not in the method, but in using the hypothesis of the thermal bath 
in all its consequences.

In a normal chemical process $\nu_1 X_1 + \nu_2 X_2 \rightleftarrows \nu_3 X_3 +\nu_4 X_4$
with atoms, molecules, nuclei or elementary particles where the total number
of particles is conserved, 
the stoichiometry of the reaction implies that 
\begin{flalign}
\frac{1}{\nu_1} \frac{d n_1}{dt}=
\frac{1}{\nu_2} \frac{d n_2}{dt}=
\frac{1}{\nu_3} \frac{d n_3}{dt}=
\frac{1}{\nu_4} \frac{d n_4}{dt}.
\label{stechio}
\end{flalign}
The four  differential equations for the concentration of each specie  are thus not independent.
Equation (\ref{stechio}) suggests that 
actually only one quantity is necessary to describe the 
changes in the number of moles/particles.
This state variable in modern thermodynamics is called 
\textit{extent of reaction}~\cite{thermo}, defined by
\begin{flalign}
d\xi=\frac{d N_1}{\nu_1}=\frac{d N_2}{\nu_2}=
\frac{d N_3}{\nu_3}=\frac{d N_4}{\nu_4}.
\label{eor_def}
\end{flalign}
At the beginning of the reaction $\xi(0)=0$. If the 
initial abundances  at $t=0$ are $N_{0,i}$,  then at each time during the reaction 
progress the abundances of all species are given by
\begin{flalign}
N_i (t)=N_{0,i}+\nu_i \xi(t),
\label{extent}
\end{flalign}
where $\xi(t)$ determined by the rate equation~\cite{thermo}
\begin{flalign}
\frac{1}{V} \frac{d\xi}{dt}=\mathcal{R}_f(N_{0,i}) -\mathcal{R}_b (N_{0,i}),
\label{rate_equation_extent}
\end{flalign}
with the initial condition $\xi(0)=0$. 
Note that in Eq.~(\ref{extent}) the rates are expressed in terms of $\xi$ and the 
initial abundances appear explicitly in the equation.

If we remove the assumption of thermal bath $n_\psi = n_{\bar{\psi}} =n_{0,\psi}$, 
the equations for WIMPs and for the bath particles would conform to that usual scheme.
As we have already discussed above, we would have only one equation, 
but  that equation would not have the standard form. 
In the creation rates the densities $n_\psi$ would appear instead of $n_{0,\psi}$,
and  we should know the initial abundance of both  WIMPs and  $\psi_i$ particles.

The presence of $dY_0 /dx$ in Eq.~(\ref{Y1eq}) should 
not hide the fact that (\ref{Y1eq})
is just another form of the algebraic equation 
(\ref{eq_bath_before_simpl}), which
accounts for the properties of the thermal bath. 

Assumption (\ref{assumption_2}) is basic for the theory of thermally produced WIMPs.
It not only constrains the creation rate, but also the extent of reaction.
Comparing  Eqs.~(\ref{rate_eq_wimps}) and (\ref{rate_eq_bath1})
with Eqs.~(\ref{eor_def}) and (\ref{rate_equation_extent}), we see that what appears in 
the left hand sides of the former is the time derivative 
of the extent of reaction, while Eq.~(\ref{extent}) tells us that $\Delta$, up to the
stoichiometric coefficient, is the same as the integrated extent of reaction.
Imagine we can stop the expansion, that is, we fix the temperature, and observe the 
evolution in time of the creation reaction $\psi\bar{\psi}\to 2\chi$ as if it were an ordinary reaction. 
Let the initial amount of both bath particles and WIMPs be $N_0$.
The maximum extent to which this reaction can proceed corresponds to the total consumption of the $\psi$, 
$\xi_\text{max}=(0-N_0)/(-1)=N_0$ where we used our convention of negative stoichiometric coefficients 
for reactants. The amount of WIMPs at the maximum hence would be $N_0+N_0=2N_0$. This
heuristically explains the Zeldovich approximate ansatz at the freeze out point.

The resulting physical picture is clear.
During the expansion WIMPs are produced by the thermal bath up to the temperature $T_\ast$
where the extent of reaction for production reaches the maximum
value. Both $T_\ast$ and the maximum depends on the WIMP mass and on the strength of interactions
between the WIMPs and the bath in a known constrained way.
Once reached this maximum WIMPs production becomes negligible, the annihilation rate reduces the abundance 
that relaxes asymptotically to a constant value because the same annihilation rate diminishes 
with further expansion/cooling.

From the mathematical point of view
the whole evolution in time/temperature of the WIMPs abundance is given by a differential-algebraic system.  
The differential equation (\ref{Yequation})  has the form $dY/dx=f(Y,Y_0,x)$,
while the  algebraic equation (\ref{Y1eq}), which can be written as 
\begin{flalign}
Y^2-Y^2_0-\frac{x^2}{C \langle \sigma_{\text{ann}} v_\text{r} \rangle}\frac{dY_0}{dx}=0,
\end{flalign}
has the form $g(Y,Y_0, d{Y}_0 /dx,x)=0$ and acts as a constraint in the interval $x\le x_\ast$.
The initial value problem $Y(x_\ast)=Y_1 (x_\ast)$ is a consistent initial value problem for the 
differential equation given the physical 
assumptions and constraints.
In facts, we have seen that the production and decoupling of WIMPs  
depends both on $dY_0/dx$ and on $Y_0$.
This is taken into account by the algebraic
equation $g(Y,Y_0, d Y_0 /dx,x)=0$.
The fact that the evolution of the abundance is differential only starting at  $x_\ast$,
is in the fact that function $f(Y,Y_0,x)$ does not depend on $dY_0 /dx$.

\section{Approximation for the relic abundance:
Revising the Zeldovich criterion}
\label{Sec_relic_approx}

The fact that the abundance at $x\le x_\ast$ is an already fixed function $Y_1(x)$
allows to find a very accurate approximate formula for the relic abundance.

Let us see what happens if we apply the freeze out approximation with $x_f = x_\ast$.
Neglecting the term $Y^2_0$ in (\ref{Y2eq}) 
and integrating with the initial condition (\ref{initialcondition}) we have 
\begin{flalign}
Y^{\text{fo}}_2(x)=
\frac{ Y_1(x_\ast)}{1+ Y_1(x_\ast) C \int_{x_\ast}^{x}
dx' 
\frac{\langle \sigma_{\text{ann}} v_\text{r}\rangle}{x'^2}} .
\label{Y2_app}
\end{flalign}
As can be seen in the bottom panel of Fig.~\ref{Fig1}, 
Eq.~(\ref{Y2_app}), the green dotted-dashed line, underestimates the exact solution $Y_2(x)$, 
red dashed line, by  20$\%$ in the freeze out zone. In facts, near $x_\ast$,
$Y_0$ is not much smaller than the actual abundance $Y_2(x)$ 
and the approximation is not so good. 

As far as concerns the analytical approximation, one is not 
obliged to use $x_\ast$, but can choose a better point.
The function $\Delta$ posses another characteristic point, where it is intersected
by the equilibrium function $Y_0$, $\Delta(x_2)=Y_0(x_2)$,
see the inlay in the top panel and the bottom panel of Figure~\ref{Fig1}.
At $x=x_2$, hence $Y_1(x_2)=2Y_0 (x_2)$ and $\delta(x_2)=1$.  
With the same numbers as before, requiring $\delta(x_2)=1$ we find the numerical value $x_2 =21.97$.

As it evident in the bottom panel, in the interval between $x_\ast$ and $x_2$,
$Y_1$ and $Y_2$ are still very close, while only after $x_2$ the true abundance $Y_2$
starts to deviate significantly from $Y_1$. This behavior suggests to use $x_2$ and 
\begin{flalign}
Y_2 (x_2)\simeq Y_1(x_2)=2Y_0(x_2),
\end{flalign}
as initial conditions in Eq.~(\ref{Y2_app}).
In the bottom panel we see the clear improvement: now the black solid line underestimates the
red dashed line  by 4$\%$ around $x_2$. At larger $x$ is practically superimposed $Y_2$,
thus asymptotically the approximation will be even better.
In facts, the relic abundance is hence given by
\begin{flalign}
Y^{\text{fo}}_2 (\infty)=\frac{ 2 Y_0 (x_2)}{1+ 2 Y_0(x_2) C 
\frac{\langle \sigma_\text{ann} v_\text{r}\rangle_0}{(n+1) x^{n+1}_2} },
\label{Y2inf}
\end{flalign}
where we have taken a power law dependence on the temperature,
$\langle  \sigma_{\text{ann}} v_\text{r}\rangle=\langle \sigma_\text{ann} v_\text{r}\rangle_0 
x^{-n}$~\cite{ScherrerTurner} and sent to infinity $x$ in (\ref{Y2_app}).
We find, for example, $Y_2(10^8)/ Y^{\text{fo}}_2 (\infty) =1.01$. 
In the case of $P$-wave scattering, $\langle 
\sigma_{\text{ann}} \text{v}_{\text{r}}\rangle =\langle \sigma_{\text{ann}} \text{v}_{\text{r}}\rangle_0/x$, 
with $\langle \sigma_{\text{ann}} \text{v}_{\text{r}}\rangle_0=10^{-10}$ GeV$^{-2}$,
the new points are $x_{f,P}=18.19$ and $x_{2,P}=19.11$. The asymptotic ratio using $x_{2,P}$ is
$Y_2(10^8)/ Y^{\text{fo}}_2 (\infty) =1.02$.
Eq.~(\ref{Y2inf}) gives the relic abundance with an accuracy at the level of 1$\%-2\%$, which is extremely good.
We emphasize that the function $Y_1(x)$ extended up to $x_2$, together with the function $Y^{\text{fo}}_2(x)$
starting at $x_2$, furnishes a semi-exact analytical description of the whole evolution of the actual WIMPs abundance.

Note that from Eq.~(\ref{delta}), the condition $\delta(x_2)=1$ is equivalent to
\begin{flalign}
-\frac{1}{Y_0} \frac{dY_0}{dx}= 
3 \frac{C}{x^2} \langle \sigma_{\text{ann}}v_\text{r}   \rangle Y_0\;\;\text{at $x=x_2$}.
\label{x2mio_general}
\end{flalign}
In the non relativistic limit as in Eqs.~(\ref{Zeld_criterion_1}), (\ref{log_der}), (\ref{popular_criterion}) it becomes 
\begin{flalign}
\frac{3\Gamma}{Hx} \sim 1  ,
\label{x2mio_nonrel}
\end{flalign} 
or, in terms of characteristic times, we have $t^{-1}_1=3\tau^{-1}$.

We can establish now the connection with the Zeldovich 
criterion~\cite{VDZ,ZOP,DolgovZeldovich}.
In Ref.~\cite{VDZ} the freeze out temperature was derived
with the criterion 
\begin{flalign}
4 \langle \sigma_{\text{ann}}v_\text{r}   \rangle n_0 t \frac{T}{m}\sim 1.
\label{VDZcriterion}
\end{flalign}
The origin of this criterion can be understood reasoning as in the Introduction.
The characteristic time of the expansion 
$t^{-1}_{1} =(1/n_0)  dn_{0}/dt$ is evaluated using the nonrelativistic
density $n_0$=cost.$(mT)^{3/2}\exp(-m/T)$ and the relations between time and 
temperature in the radiation era, $T\propto t^{-1/2}$ and $dT/dt=(1/2)T/t$. 
Changing variable from $t$ to $T$,
and taking the derivative of $n_0$ neglecting the variation of the term $T^{3/2}$,
we have $t_{1}\simeq 2t T/m$. 

The characteristic time of variation
of the abundance due to interactions can be estimated from the kinetic equation written 
as $dY/dt=-\langle \sigma_{\text{ann}}v_\text{r} \rangle s (Y^2-Y^2_0)$.
Near equilibrium, in the right hand side we can approximate $s (Y^2-Y^2_0)\approx 2n_0 
(Y-Y_0) $,
giving a characteristic time of variation 
$\tau=1/(2\langle\sigma_{\text{ann}}v_\text{r} \rangle n_0)$.
By requiring $t_1=\tau$ we obtain Eq.~(\ref{VDZcriterion}).

We can write Eq.~(\ref{VDZcriterion}) in a form analogous to 
(\ref{popular_criterion}) and (\ref{x2mio_nonrel}) by using
the relation $t=1/(2H)$,
\begin{flalign}
\frac{2\Gamma}{Hx} \sim 1,
\label{VDZcriterion_mio}
\end{flalign}
or, more generally, like (\ref{Zeld_criterion_1}) and (\ref{x2mio_general}),
\begin{flalign}
-\frac{1}{Y_0} \frac{dY_0}{dx}= 
2 \frac{C}{x^2} \langle \sigma_{\text{ann}}v_\text{r}   \rangle Y_0\;\;\text{at $x=x_{VDZ}$}.
\label{VDZcriterion_mio_2}
\end{flalign}
In terms of the Scherrer-Turner ansatz, 
this criterion corresponds to $c(c+2)=2$,
that is $c=\sqrt{3}-1$ 
and $Y(x_{VDZ})=\sqrt{3}Y_0(x_{VDZ})$.
Using Eq.~(\ref{VDZcriterion_mio_2}), and the previous numbers,
we find $x_{DVZ}=21.57$. Using $\sqrt{3}Y_0(x_{DVZ})$ in Eq.~(\ref{Y2inf})
instead of $2Y_0(x_2)$, we find that the ratio 
$Y_2(10^8)/Y^{\text{fo}}_2 (\infty) =1.014$ for the $S$-wave case.
A similar calculation gives $1.028$ for $P$-wave case.

The criterion thus gives very good approximation for the asymptotic value
when compared with the numerical integration, as stated in Ref.~\cite{VDZ},
just a little bit worse than our proposed criterion. This is clear from the
bottom panel of Fig.~\ref{Fig1}. Selecting a point between $x_*$ and $x_2$,
gives a freeze out approximation curve that lays between the dot-dashed and 
the full-black line: the point  $x_{VDZ}$ is near $x_2$, thus the freeze 
approximation curve starting from there will approximate well 
the exact numerical solution given by the dashed line.

\section{chemical potential of WIMPs and entropy production}

We have seen that at the initial instant  all chemical potentials are zero and chemical equilibrium holds.
With the expansion/cooling of the universe and the
bath particles with $\mu_{\psi_i} =0$, the WIMPs start to develop a nonzero chemical potential.

A non equilibrium chemical transformation is characterized by the state variable called \textit{affinity}, 
which is defined as~\cite{thermo} 
\begin{flalign}
\mathcal{A}=-\sum\limits_{i}\nu_i \mu_i,
\label{affinity_def}
\end{flalign}  
where $\mu_i$ are the chemical potentials of the species.
The affinity is zero in chemical equilibrium, $\mathcal{A}_{\text{eq}}=0$, and is related to the rates by~\cite{thermo}
\begin{flalign}
\mathcal{A}= T \ln \left(\frac{\mathcal{R}_f} {\mathcal{R}_b}\right).
\label{affinity_rates}
\end{flalign}
A  nonzero affinity drives the process out of chemical equilibrium.
From Eq.~(\ref{affinity_def}), with the convention that stoichiometric coefficients are negative for reactants, we have 
\begin{flalign}
\mathcal{A}=2 \mu_\chi.
\label{affinity_chi}
\end{flalign}
Since the explicit form of the rates is known,
\begin{flalign}
\mathcal{R}_f =\frac{\langle \sigma_{\text{ann}} v_\text{r}\rangle}{2} s^2 Y^2,
\;\;\;\;\;
\mathcal{R}_b = \frac{\langle \sigma_{\text{ann}} v_\text{r}\rangle}{2} s^2 Y^2_0,
\label{f_b_rates_Y}
\end{flalign}
using  Eq.~(\ref{affinity_rates}) we find 
\begin{flalign}
\mathcal{A}=2\frac{m}{x}\ln\left(\frac{Y}{Y_0}\right),
\label{affinity_Y}
\end{flalign}
and from Eqs.~(\ref{affinity_chi}) and (\ref{affinity_Y}) the chemical potential is
\begin{flalign}
\mu_\chi = \frac{m}{x}\ln\left(\frac{Y}{Y_0}\right).
\label{mu_Y}
\end{flalign}
Here $Y$ is the piecewise abundance (\ref{piecewiseY}).
\begin{figure}[t!]
\includegraphics*[scale=0.36]{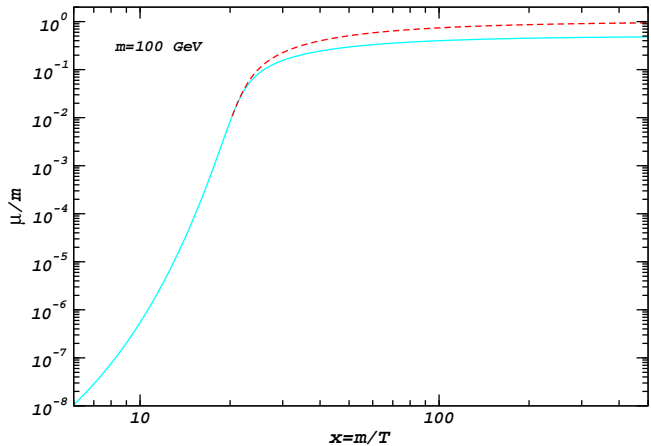}
\caption{
Evolution of the WIMPs chemical potential given by
Eq.~(\ref{mu_Y}) with the same numerical inputs of Fig.~\ref{Fig1}. 
The abundance $Y$ is the 
piecewise function (\ref{piecewiseY}). The solid line 
corresponds to $Y_1$ and the dashed line to $Y_2$ as in
Fig.~\ref{Fig1}. 
}
\label{Fig2}
\end{figure}

At small temperatures, large $x$, when the rates are virtually zero and the
WIMPs abundance becomes constant, $\mu_\chi$ cannot grow arbitrarily and must tend to an asymptotic value.
The chemical potential being the  free energy per particle that can released or gained when particles 
are produced or annihilated, the asymptotic value is the mass of the WIMP:
\begin{flalign}
\lim\limits_{x\to\infty}\mu_\chi (x)=m_\chi.
\end{flalign}
In fact, at large $x$, $\ln(Y_2/Y_0)\sim \ln(e^{x})\sim x$, 
thus from Eq.~(\ref{mu_Y}) we have $\mu_\chi \sim m$. 
Instead, $\ln(Y_1/Y_0)= \ln(1+\delta(x))\sim\ln( e^{x/2})\sim x/2$ 
thus the chemical potential would tend to $m_\chi/2$. 
This is consistent with the fact $Y_1$ gives the abundance only up to  $x_f$
while for $x>x_f$ the evolution  is given by $Y_2$.
The behavior is clearly seen in Fig.~\ref{Fig2}.

The rate equation for the extent of reaction can also be written in terms of the affinity as~\cite{thermo} 
\begin{flalign}
\frac{1}{V} \frac{d\xi}{dt}=\mathcal{R}_f (1-e^{-\mathcal{A}/T}).
\label{rate_eq_affinity}
\end{flalign}
As a consistence check, note that 
$(1/V) d\xi/dt=-{1}/{2} ({1}/{a^3}) d(n a^3)/dt$.
From Eq.~(\ref{affinity_Y}) it follows $1-\exp(-\mathcal{A}/T)=1-Y^2_0/Y^2=1-n^2_0/n^2$
and $\mathcal{R}_f =(\langle \sigma_{\text{ann}} v_\text{r}\rangle/ 2) n^2$,
thus Eq.~(\ref{rate_eq_affinity}) gives the standard equation.

The freeze out is a  nonequilibrium  irreversible process that increases the
entropy of the Universe. We can easily verify that anyway the produced entropy is negligible
compared with the entropy of the radiation.
The entropy production in chemical reactions is determined
by affinity and the extent of reaction through~\cite{thermo}
\begin{flalign}
\frac{1}{V} \frac{dS_r}{dt}=\frac{1}{V}\frac{\mathcal{A}}{T}\frac{d\xi}{dt}=(\mathcal{R}_f -\mathcal{R}_b )\ln (\frac{\mathcal{R}_f} {\mathcal{R}_b}).
\label{entropy_prod}
\end{flalign}
This quantity is always positive as required by the second law of thermodynamics. 
By changing variable from time to $x$ we find
\begin{flalign}
\frac{1}{S}\frac{dS_r}{dx}= \frac{C}{x^2} {\langle \sigma_{\text{ann}} \text{v$_\text{r}$}\rangle}(Y^2-Y^2_{0})\ln (\frac{Y}{Y_{0}}),
\end{flalign}
where $S_r$ is the entropy  produced in the reaction and $S=s a^3$ is cosmic entropy density.
It is easily seen that the scale of $S_r$ relative to $S$ is set by the magnitude of $\langle \sigma_{\text{ann}} \text{v$_\text{r}$}\rangle$. In our numerical example hence $(dS_r/dx)/S\simeq 10^{-10}$:
the entropy produced in the production and decoupling of WIMPs is thus completely negligible
compared with $S$, which thus remains constant during the process.

\end{document}